\documentclass[a4paper,twoside,onecolumn,12pt]{article}

\usepackage{t1enc}
\usepackage{times}
\usepackage[cp852]{inputenc}
\usepackage[english]{babel}
\usepackage{graphicx}
\usepackage{fancyhdr}
\usepackage{amsmath}
\usepackage{upgreek}
\def\thebibliography#1{\section*{References}\list
 {}{\setlength\labelwidth{1.4em}\leftmargin\labelwidth
 \setlength\parsep{0pt}\setlength\itemsep{0pt}
 \setlength{\itemindent}{-\leftmargin}
 \usecounter{enumi}}
 \def\newblock{\hskip .11em plus .33em minus -.07em}
 \sloppy
 \sfcode`\.=1000\relax}

\newcommand{\f}{\frac}

\newcommand{\bfr}{\boldsymbol{r}}
\newcommand{\bfa}{\boldsymbol{a}}

\newcommand{\bc}{\begin{center}}
\newcommand{\be}{\begin{equation}}
\newcommand{\ee}{\end{equation}}
\newcommand{\ec}{\end{center}}

\newcommand{\m}{\mathrm}

\newcommand{\spose}[1]{\hbox to 0pt{#1\hss}}
\newcommand{\lta}{\mathrel{\spose{\lower 3pt\hbox{$\mathchar"218$}}
 \raise 2.0pt\hbox{$\mathchar"13C$}}}
\newcommand{\gta}{\mathrel{\spose{\lower 3pt\hbox{$\mathchar"218$}}
 \raise 2.0pt\hbox{$\mathchar"13E$}}}

\def\@cite#1#2{{#1\if@tempswa , #2\fi}}
\def\@biblabel#1{}

\def\utw{\smash{\rlap{\lower5pt\hbox{$\sim$}}}}
\def\udtw{\smash{\rlap{\lower6pt\hbox{$\approx$}}}}

\begin{document}

%---------------------------------------------------
%
%        Definiujemy nag│ˇwki i stopki
%
%---------------------------------------------------
\fancyhead{} \fancyhead[CO]{Hot Jupiters and Central Cavities...}
\fancyhead[CE]{S. Starczewski, A. J. Gawryszczak, R. W\"unsch \&
M. R\'o\.zyczka} \fancyhead[LE,RO]{\thepage} \fancyfoot{}
\renewcommand{\headrulewidth}{0.4pt}
\renewcommand{\footrulewidth}{0pt}

\voffset=-1cm
\marginparwidth=0pt
\oddsidemargin=0.3cm
\evensidemargin=-0.3cm

\title{Hot Jupiters and Central Cavities of Protoplanetary Discs}
\author{Szymon Starczewski$^1$, Artur J. Gawryszczak$^1$,\\ Richard
W\"unsch$^{1,2}$ \& Micha\l\ R\'o\.zyczka$^1$ \\ $^1$N. Copernicus
Astronomical Center, Warsaw, Poland \\ $^2$Astronomical Institute,
Academy of Sciences of the Czech Republic, \\ Praha, Czech
Republic \\ e-mail: (star, gawrysz, mnr)@camk.edu.pl \\
richard.wunsch@matfyz.cz}
\date{}
\maketitle

\begin{abstract}
We investigate numerically the orbital evolution of massive
extrasolar planets within central cavities of their parent
protoplanetary discs. Assuming that they arrive at the inner edge
of the disc due to type II migration, we show that they spiral
further in. We find that in magnetospheric cavities more massive
planets stop migrating at a larger distance from the edge of the
disc. This effect may qualitatively explain the correlation
between masses and orbital periods found for massive planets with
$P$ shorter than 5 days.

{\bf key words:} extrasolar planets, planet formation, protoplanetary discs
\end{abstract}
\section{Introduction} \label{sect:_spp_orig}
The standard model of giant planet formation through
core-accretion and envelope capture assumes that a solid core is
formed first by accretion of planetesimals, and when it becomes
massive enough it begins to accrete gas from the surrounding
nebula. It is generally believed that gas giants form exterior to
the snow line, i.e. where the disc is cold enough for water to
condense, thus increasing the amount of solids available for core
buildup (in very massive discs the cores can reach the critical
mass also interior to the snowline). Once a sufficiently massive
planet is formed, it opens a gap in the disc and undergoes
migration of type II; see e.g. Ida \& Lin (2004) and references
therein. Less massive planets, unable to open a gap, are likely to
undergo migration of type I; see e.g. Ward (1997). In either case,
the final orbit of the planet is expected to be much tighter than
the original one.

\begin{figure}
 \includegraphics{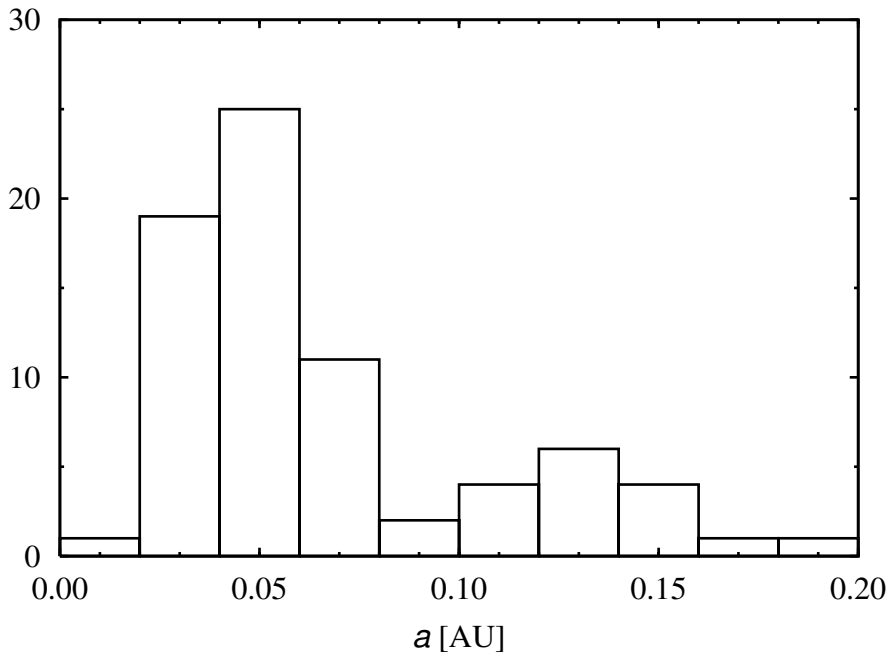}
 \caption {The histogram of semimajor axes of Hot Jupiters. Based
 on data from Schneider (2007).}
 \label{fig: hist_a}
\end{figure}

The prime candidates for planets that have undergone extensive
migration are objects on the tightest orbits, whose orbital radii 
are clustered between $a \approx 0.03$ a.u. and $a \approx 0.06$ a.u.  
(Fig. \ref{fig: hist_a}). A major fraction of these objects are the 
so-called Hot Jupiters (hereafter HJ), i.e. giant gaseous planets 
of Jupiter or Saturn type. Several ideas have been proposed how
to stop the migration at small orbital radii, including presence
of a central cavity in the disc, tidal interactions with the star,
and mass loss from the planet due to the Roche lobe overflow
(Trilling et al. 1998). The cavity may originate due to the
truncation the disc by the magnetosphere of the star (Lin,
Bodenheimer \& Richardson 1996; hereafter LBR) or the
magnetorotational instability, which accelerates the accretion
flow, leading to a strong drop in surface density up to the radius
where the temperature falls below 1500 K (Kuchner and Lecar 2002).

The magnetospheric truncation hypothesis is strongly supported by
the fact that T Tau stars have magnetic fields with intensities up
to $\sim4.5\times10^3$ G (Symington et al. 2005). For typical
accretion rates inferred for these objects ($10^{-9}$ to $10^{-7}
M_\odot$ yr$^{-1}$) the disc should be truncated at a distance of
a few $R_\star$ (stellar radii) from the star. This prediction
finds some observational support, as circumstellar cavities of a
few $R_\star$ are required to explain the widths of CO lines most
likely originating in a gas on Keplerian orbits around T Tau stars
(Bouvier et al. 2006 and references therein). An obvious
conclusion is that many (perhaps most) HJs with semimajor axes
smaller than $\sim0.06$ AU must have spent some time {\it within
an active magnetosphere}, i.e. within the central cavity of their
parent protoplanetary disc.

The orbital evolution of a planet within the magnetospheric cavity
was considered by LBR. In their scenario, the disc is truncated at
the magnetospheric radius $r_\m{m}\approx0.08$ AU. The star
corotates with the inner edge of the disc, to which it is
magnetically locked. A planet that has migrated into the cavity
experiences negative torques from both the disc and the star.
However, once it has spiralled past the 2:1 resonance at the
orbital radius $a_{2:1}=0.63r_\m{m}$, the torque from the disc
becomes drastically weaker. The stellar torque also weakens in
time due to the contraction of the star on its way to the main
sequence (in fact, it may even reverse the sign if the angular
momentum of the star is conserved). As a result, LBR expect the
migration to effectively stop at ~$\sim0.05$ AU. The same basic
scenario is repeated in many later papers, e.g. Lin et al. (2000),
Kuchner \& Lecar (2002), Eisner et al. (2005), Santos et al.
(2005) or Romanova \& Lovelace (2006). However, despite of its
popularity, it has never been verified by numerical simulations.

LBR and their followers implicitly assume that i) the
magnetosphere simply truncates the disc at $r_\m{m}$, leaving it
undisturbed for $r>r_\m{m}$, and ii) the gravity of the matter
accreting onto the star along the field lines and/or blown away
from the star as a magnetized wind can be neglected. At a first
glance, neither of these assumptions looks plausible; however they
are quite strongly supported by several arguments. First, Romanova
\& Lovelace (2006) demonstrate that in many cases the
magnetospheric cavity is indeed almost empty. Second, the
simulations of Long, Romanova \& Lovelace (2005; hereafter LRL)
show that in many cases the gravity of the wind is indeed
negligible (typical of their type II models is an outflowing
corona with a density $\sim 10^3$ times lower than the midplane
density of the disc). Third, in some LRL models the accreting
matter forms a nearly spherical thin shell with narrow polar
funnels. If the shell is nearly uniform, then the gravitational
potential it generates within the magnetosphere is nearly
constant, and the gravitational effect of the accretion flow on
the planet residing in the magnetosphere may be rather weak. Of
course, because of numerous plasma instabilities, a smooth
magnetospheric shell is an oversimplification. However, it is
quite feasible that the time-averaged contribution to the torque
on the planet from the fluctuating part of the density field is
insignificant.

Based on these arguments, the magnetospheric migration can be
followed by simplified simulations in which i) the accretion disc
has a sharp inner edge at the magnetospheric radius $r_\m{m}$, ii)
the matter that arrives at $r_\m{m}$ does not enter the
magnetosphere but is lifted off the disc, and iii) at $r<r_\m{m}$
the net gravitational effect of the magnetospheric flow is
neglected. Following the implicit assumption made by LBR, we also
assume that beyond $r_\m{m}$ the disc is practically undisturbed
by magnetic fields, i.e. that non-Keplerian effects in its
dynamics are dominated by the gravity of the planet rather than
Lorentz force. The latter assumption, whose validity we check {\it
a posteriori} in Sect. \ref{sect:_results}, allows us to calculate
the torque from the disc on the planet with the help of standard
two-dimensional simulations.

The simulations are presented in Sect. 2. In Sect. 3 we discuss
their results, and speculate about the origin of the semimajor axes
distribution shown in Fig. \ref{fig: hist_a}. The paper is
supplemented by an appendix containing a report on the code
calibration test.
\section{Numerical simulations}\label{sect:_simulations}
We employ the FLASH code, which operates on a {\it
block-structured} grid (Fryxell et al. 2000). Within each block
the equations of hydrodynamics are solved using a conservative
third-order Eulerian scheme (Colella \& Woodward 1984). FLASH can
build a hierarchy of progressively finer grids by "halving" the
blocks in each coordinate whenever refinement is required. For the
present simulations, however, a uniform (nonrefined) grid proved
to be sufficient.
\subsection{Problem setup}
We consider a system composed of a central star with mass
$M_\star$ (in the simulations $M_\star = 1 M_\odot$), a planet
with mass $M_\m{p}$, and a thin non-selfgravitating disc truncated
at a distance $r_\m{m}$ from the star. The disc is locally
isothermal, and the vertically integrated pressure $p$ is related
to the surface density $\Sigma$ through
\be
  p=\Sigma c_{\m{s}}^{2}\,,
  \label{eq:eos}
\ee
 with the local sound speed given by
\be
  c_{\m{s}}= H\Omega_\m{K}\,,
  \label{eq:sounds}
\ee
 where $\Omega_\m{K}=\sqrt{\m{G}M/r^3}$ is the Keplerian angular
velocity, and $H$ is the half-thickness of the disc. The aspect
ratio $h\equiv H/r$ is assumed constant over $r$ and equal to
0.05.

We do not introduce any explicit magnetic field. Following the
arguments presented in Sect. \ref{sect:_spp_orig}, we simply
assume that for $r< r_\m{m}$ the accretion flow does not influence the
planet in any way. With such an assumption, in a coordinate system
centered on the star the orbital evolution of the planet is
described by the standard equation
\be
  \frac{{\rm d}^{2}{\bfa}}{{\rm d}t}=
  -\frac{\m{G}(M_\star+M_\m{p})}{{a}^{3}}{\bfa}-\nabla\Phi_\m{d},
  \label{eq:peom}
\ee
 where $\bfa$ is the position vector of the planet, and
$\Phi_\m{d}$ is the gravitational potential of the disc. The formula
for the potential,
\be
  \Phi_\m{d}=-\m{G}\int_{S}\frac{\Sigma(\bfr)}{|\bfr-{\bfa}|}
  {\rm d}\bfr+\m{G}\int_{S}\frac{\Sigma(\bfr)}{r^{3}}{\bfa}\cdot\bfr
  {\rm d}\bfr\,,
  \label{eq:dgp}
\ee
 contains the indirect term accounting for the non-inertiality of the
coordinate system, and the integration in (\ref{eq:dgp}) is
performed over the surface of the disc.

The magnetospheric radius is given by the standard formula
\be
  r_\m{m}=\eta\left(\frac{B_\star^4R_\star^{12}}{\m{G}M_\star\dot{M}^2}
  \right)^{1/7}\,,
  \label{eq:rm}
\ee
 where $B_\star$, $R_\star$, $M_\star$, $\dot{M}$ and $\eta$ are,
respectively, field strength at the surface of the star, stellar
radius, stellar mass, accretion rate, and a dimensionless factor
of order unity. For a star with an aligned dipole field
$0.5\le\eta\le1.0$; see e.g. Lai (1999) and references therein.
Normalized to the standard parameters of T Tau stars, equation
(\ref{eq:rm}) reads
\be
  \frac{r_\m{m}}{{R}_\odot}
  =4.29\eta\left[\frac{\left(\frac{B_\star}{1000\,\m{Gs}}\right)^4
                      \left(\frac{R_\star}{{R}_\odot}\right)^{12}}
  {\frac{M_\star}{{M}_\odot}
  \left(\frac{\dot{M}}{10^{-8}{M}_\odot/\m{yr}}\right)^2}
  \right]^{1/7}\,,
  \label{eq:rmn}
\ee
 We take $r_\m{m} = 12 R_\odot = 0.056$ AU, which agrees with the
distance range suggested by the distribution shown in Fig.
\ref{fig: hist_a} (note also that for the standard values of
$B_\star$, $M_\star$ and $\dot{M}$ the adopted value of $r_\m{m}$
corresponds to an entirely reasonable range of stellar radii
%$2R_\odot\stackrel{<}{\sim}R_\star\stackrel{<}{\sim}3R_\odot$).
$2R_\odot \leq R_\star \leq 3R_\odot$).
\subsection{Results}\label{sect:_results}
The simulations were performed on a polar grid of $n_{r}=512$ and
$n_{\phi}=128$ points, extending from $r_\m{in}=r_\m{m}$ to
$r_\m{out}=0.356$ AU. We checked that doubling $n_{r}$ and
$n_{\phi}$ changes the value of the disc torque on the planet by
less than 2\%, which proved that our basic resolution was
sufficient for the present problem. At $r_\m{in}$ a free outflow
boundary condition was applied, while at the outer edge of the
disc we implemented the prescription proposed by de Val-Borro et
al. (2006) to damp reflections from the boundary of the grid.
Specifically, after each time-step the equation
\be
  \f{{\rm d}X}{{\rm d}t}=-\f{X-X_{0}}{P}R(r)
  \label{eq:obc}
\ee
 was solved, where $X$ stands for $\Sigma$ or velocity component,
$X_{0}$ is the initial profile of a particular variable, and
$R(r)$ is a parabolic ramp function which decreases smoothly from
$1$ at $r=r_\m{out}$ to $0$ at $r=0.84r_\m{out}$. The default FLASH
artificial viscosity with $c_{\m{visc}} = 0.1$ was used to
stabilize the solution.

Initially the disc is uniform, and $a = 0.062$ AU. To relax the
model, we keep the planet on the initial orbit until it carves a
quasistationary gap in the disc, exterior to which a spiral
density pattern emerges. When the relaxation procedure is
completed (usually after $\sim100$ orbital periods of the planet),
the time-counter is reset to 0, and the planet begins to evolve
according to equation (\ref{eq:peom}) which we integrate with a
fourth-order Runge-Kutta scheme. We performed simulations for two
values of the mass of the planet ($M_\m{p}$ = 0.5$M_\m{J}$ and
$M_\m{p}$ = 1$M_\m{J}$, where $M_\m{J}$ is the mass of Jupiter),
and two values of the mass of the disc contained in the grid
($M_\m{d}$ = 10$M_\m{J}$ and $M_\m{d}$ = 100$M_\m{J}$). The latter
were chosen unrealistically large in order to speed the evolution
up, so that noticeable effects could be produced in a reasonable
CPU time. Within the adopted scenario the gravitational torque $T$
from the disc on the planet should scale linearly with $M_\m{d}$,
and this is what we observe in our simulations. Thus, the fact
that our discs are far too massive does not influence the
conclusions of the paper.

\begin{figure}[!t]
  \includegraphics[angle=-90,width=\textwidth]{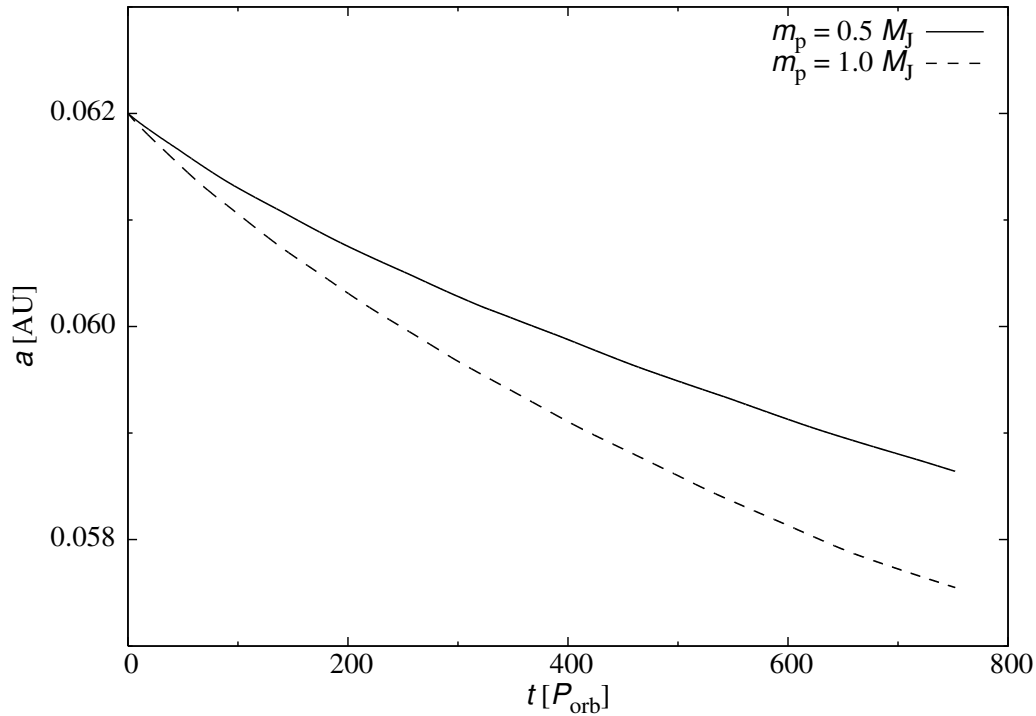}
  \caption
  {\normalsize\label{mig10}
  The migration of planets in a disc with $M_\m{d}$ = 10$M_\m{J}$. Solid:
  $M_\m{p}$ = 0.5$M_\m{J}$. Dashed: $M_\m{p}$ = 1$M_\m{J}$.}
\end{figure}
\begin{figure}[!t]
  \includegraphics[angle=-90,width=\textwidth]{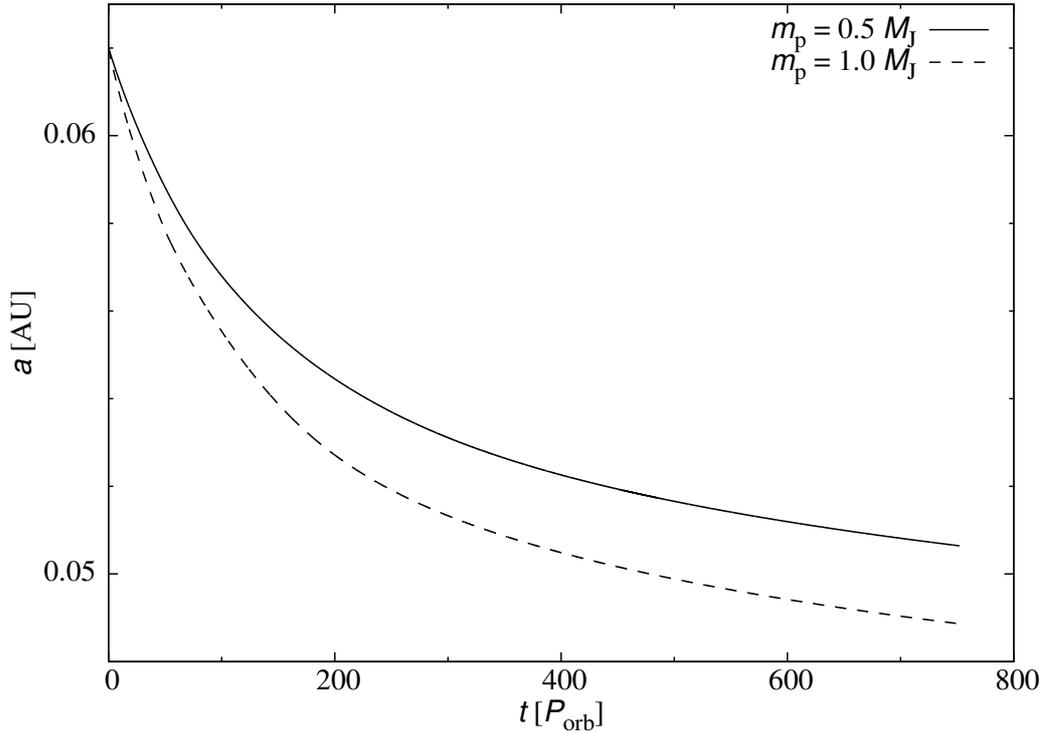}
  \caption
  {\normalsize\label{mig100}
  As in Fig. \ref{mig10}, but in a disc with $M_\m{d}$ = 100$M_\m{J}$
  }
\end{figure}
As soon as the planet has been released from its initial orbit, it
begins to migrate into the magnetosphere. The migration process is
illustrated in Figs. \ref{mig10} and \ref{mig100}  for discs with
$M_\m{d}$ = 10$M_\m{J}$ and $M_\m{d}$ = 100$M_\m{J}$,
respectively. As expected, a planet with given $M_\m{p}$ migrates
faster in more massive discs. Figs. \ref{mig10} and \ref{mig100}
also demonstrate that in a disc with given $M_\m{d}$ the migration
rate is faster for more massive planets.

\begin{figure}[!t]
  \includegraphics[angle=-90,width=\textwidth]{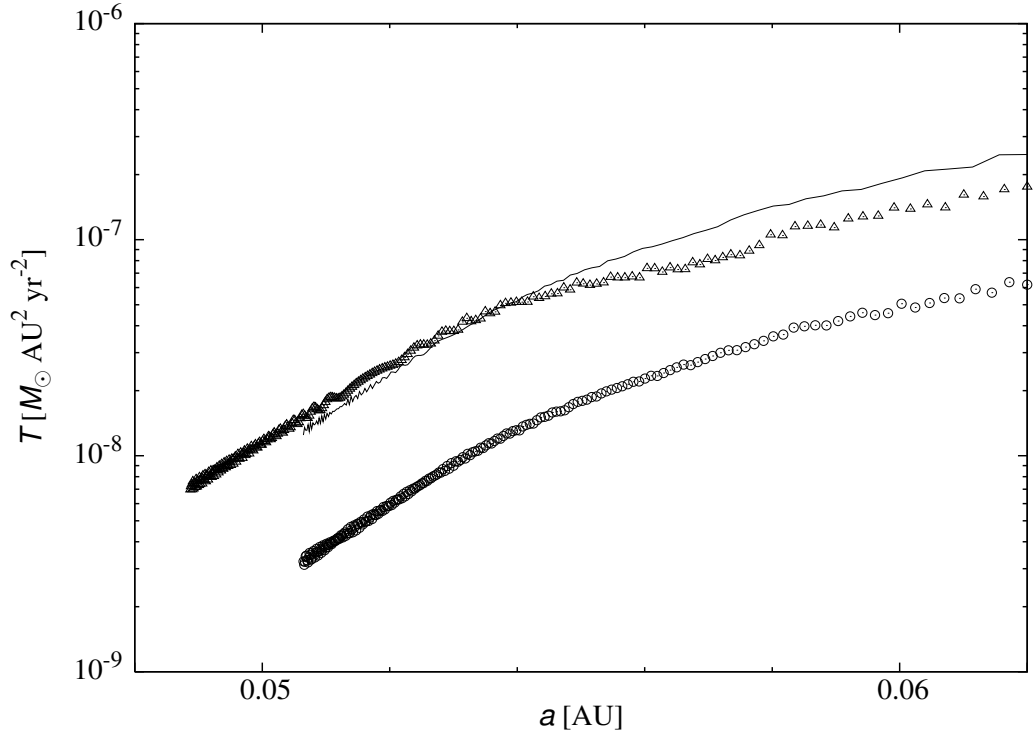}
  \caption
  {\normalsize \label{torq_d1}
  The absolute value of the disc torque on the planet as a function of the
  distance from the central star. Triangles: $M_\m{p}$ = 1$M_\m{J}$.
  Circles: $M_\m{p}$ = 0.5$M_\m{J}$. Thin line: torques for $M_\m{p}$ = 0.5$M_\m{J}$
  multiplied by 4. Vertical axis: values obtained from the simulation
  with $M_\m{d}$ = 100$M_\m{J}$, scaled to $M_\m{d}$ = 0.2 $M_\m{J}$.
  }
\end{figure}
Finally, in Fig. \ref{torq_d1} we plot $|T|$ as a function of $r$
for both values of $M_\m{p}$ (note that after an initial
adjustment the torque scales with $M_\m{p}^2$). In the following
we argue that for realistic values of $M_\m{d,p}$ the
gravitational torque is comparable to or weaker than the magnetic
torque $T_\m{m}$ on the disc from the star. In the following, $T$
and $T_\m{m}$ will stand for absolute values of gravitational and
magnetic torque, respectively.

The centrally peaked minimum mass solar nebula (MMSN) with the
surface density profile
\be
  \Sigma_1=10^3\mathrm{g}\,\mathrm{cm}^{-2}\left(\frac{r}{\mathrm{AU}}
  \right)^{-3/2}
  \label{eq:mmsn}
\ee
 (Ruden 1999), and the more gently peaking disc simulated by
Nelson et al. (2000), with the profile
\be
  \Sigma_2=10^3\mathrm{g\,cm}^{-2}\left(\frac{r}{\mathrm{AU}}
  \right)^{-1/2}\,
  \label{eq:nelson}
\ee
 yield, correspondingly, $M_\m{d}\approx1.5 M_\m{J}$ and
$M_\m{d}\approx0.03 M_\m{J}$. For further estimates we take the
geometrical mean of these two values, i.e. $M_\m{d}=0.2 M_\m{J}$.
The standard formula for $T_\m{m}$
\be
  T_\m{m}=\frac{B_\star^2R_\star^6}{r_\m{m}^3}
  \label{eq:tm}
\ee
 (Lai 1999) contains weakly constrained parameters $B_\star$ and
$R_\star$. However, using (\ref{eq:rm}) with $\eta=1$ we see that
the characteristic time scale
\be
  \tau_\m{m}\equiv \frac{J_\m{p}(r_\m{m})}{T_\m{m}}
  =\frac{M_\m{p}\sqrt{GM_\star r_\m{m}}\,r_\m{m}^3}{B_\star^2R_\star^6}
  =\frac{M_\m{p}}{\dot{M}}
  \label{eq:taum}
\ee
 (where $J_\m{p}(r_\m{m})$ is the orbital angular momentum of
the planet at the edge of the magnetosphere), does not depend on
$B_\star$ or $R_\star$. For $M_\m{p}=1M_\m{J}$ and
$\dot{M}=10^{-8}M_\odot$ ${\m{yr}}^{-1}$ equation (\ref{eq:taum})
yields $\tau_\m{m}=10^5$ yr. This is the time scale on which a
planet placed at $r_\m{m}$ would migrate if the torque from the
disc was equal to $T_\m{m}$. We shall compare it with the
timescale of migration due to the gravitational torque,
$\tau_\m{g}\equiv J_\m{p}(r_\m{m})/T$.

The orbital angular momentum of a 1$M_\m{J}$ planet placed at
$r_\m{m}=0.056$ AU is equal to $1.5\times10^{-3}
M_\odot$AU$^2$yr$^{-1}$. From Fig. \ref{torq_d1} we have
$T(r_\m{m})= 7.5\times10^{-8} M_\odot$AU$^2$yr$^{-2}$, yielding
$\tau_\m{g} = 2\times10^4$ yr. Thus, $T=5T_\m{m}$, and the
assumptions of Sect. \ref{sect:_spp_orig} are satisfied. However,
already when the planet has migrated to $a=0.9r_\m{m}=0.05$ AU,
$T$ drops to $10^{-8} M_\odot$AU$^2$yr$^{-2}$, and $\tau_\m{g}$
increases to $1.5\times10^5$ yr. This means that $T_\m{m}$ is now
larger than $T$, i.e. the inner edge of the disc is more strongly
influenced by MHD effects rather than gravity. For the $0.5
M_\m{J}$ planet the situation is even worse, since $T\approx
T_\m{m}$ already at $r_\m{m}$. As the disc dominated by chaotic
MHD effects cannot exert a consistent negative torque on the
planet, we conclude that planets with $M_\m{p}\leq 0.5M_\m{J}$
are unable to spiral into the magnetosphere, whereas Jupiter-size
planets stop spiralling at a distance $\sim0.006$ AU from the
inner edge of the disc.
\section{Discussion}
Based on the data from the beginning of September 2004, Mazeh,
Zucker and Pont (2005); hereafter MZP, and Gaudi et al. (2005)
found that among 6 then-known transiting HJs more massive planets
tended to have shorter periods. The same effect was visible in the
whole population of 17 then-known planets with orbital periods
shorter than 5 days (hereafter: SPP), albeit with large scatter.
Presently (end of May 2007) 46 SPP are known (47$^\m{th}$ object
with $P<5^\m{d}$ is a brown dwarf HD41004 B b), 20 of which are
transiting (Schneider 2007), and it is interesting to see whether
the original correlation still persists.

\begin{figure}[!t]
 \includegraphics{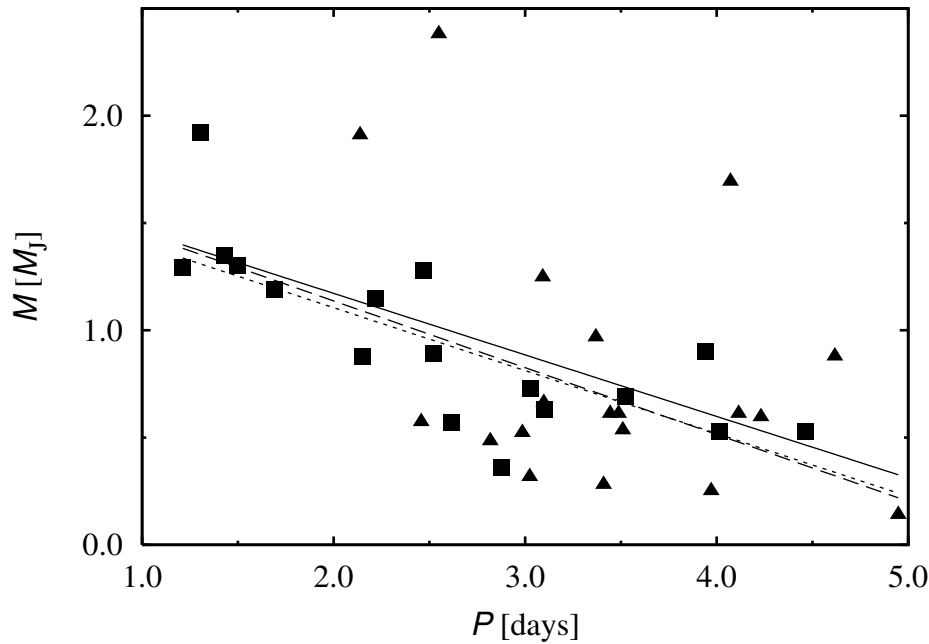}
 \caption {The mass-period relation for planets with periods
 shorter than 5 days. Based on data from Schneider (2007).
 Squares: transiting planets. Triangles:
 planets for which spectroscopic data are only available (as in MZP,
 their lower mass limits  are divided by $\uppi/4$, the expected
 value of $\sin i$). Dotted, dashed and solid
 line: fits to, correspondingly, 6 transiting planets of MZP, 17
 transiting planets with reliably determined masses known in
 January 2007, and all SPP known at the end of May 2007. }
\label{fig: m-P_diagram}
\end{figure}

Following MZP, we exclude planets in known binary systems ($\tau$
Boo b~ and HD 188753A b) as well as "Hot Neptunes" (GJ 436 b, 
55 Cnc e , Gliese 876 d, GJ 674 b and HD 219828 b), which, as MZP 
write, ``probably are of a different nature and have a different 
formation and evolutionary history''. We also exclude three transiting 
objects: SWEEPS-04, whose RV amplitude is insignificant, 
SWEEPS-11, which seems to be a brown dwarf rather than a planet 
(Sahu et al. 2006), and XO-3, whose mass is at the lower limit of 
the brown-dwarf range, and whose orbit is highly eccentric. Amazingly,
the slope of the linear fit to the $M_\m{p}(P)$ relation found for the 
14 transiting planets with reliably determined masses is nearly the 
same as the one found by MZP (Fig.\ref{fig: m-P_diagram}). Moreover,
the correlation is also visible for the whole sample of SPP, and 
the slope also nearly the same.
MZP could not provide any explanation for the observed effect, and
suggested that such an explanation should be worked out when the
$M_\m{p}-P$ correlation is better established. As the new data
confirm their findings, the effect seems to be ripe for
consideration.

Because masses of stars orbited by SPP are rather tightly
clustered around 1 $M_\odot$, whereas masses of SPP are at least
several hundred times smaller, periods of SPP are primarily
determined by their orbital radii. As a result, the $M_\m{p}-P$
correlation implies a similar connection between masses and
orbital radii, which is indeed observed (Fig. \ref{fig:
m-a_diagram}). We have shown that the "classical" scenario of the
orbital evolution inside the magnetosphere, in which planets stop
migrating at $\sim0.63r_\m{m}$, is not true. Moreover, even if it
were correct, it would not be able to account for the observed
effects. To see it, imagine an evolving population of planets
represented by points scattered on the $(a,M_\m{p})$ plane.
Initially, every planet resides at its $r_\m{m}$. We do not know
the distribution of $r_\m{m}$, but it seems reasonable to assume
that it is flat between some $r_\m{m,min}<0.056$ AU and
$r_\m{m,max}\approx0.056$ AU. If this is so, then 1) at $t=0$
there's no correlation between $M_\m{p}$ and $a$, and 2) such
correlation cannot emerge for any $t>0$, because every point is
displaced horizontally by an amount which depends on $r_\m{m}$,
but does not depend on $M_\m{p}$.

If the cavities are not of magnetospheric origin, but originate
from increased viscosity and accelerated accretion flow (Kuchner
and Lecar 2002), we encounter another problem. Although in this
case the disc is not perturbed by the magnetic field of the star,
its inner part may be strongly unstable, causing the edge of the
cavity to irregularly or semi-regularly change its location
(W\"unsch et al. 2006). As a result, any correletion between
$M_\m{p}$ and $a$ or $P$ is highly unlikely.

The results of our simulations imply however that more massive
planets tend to stop deeper in the magnetosphere, and in the
following we argue that this effect can at least qualitatively
account for the observed correlations. If the disc conforms to the
simple $\alpha$-model, then the velocity of the accretion flow is
given by
\be
 v_\m{r}=\alpha h^2 r \Omega_\m{K},
 \label{eq:_vrad}
\ee
 and, provided that $\alpha$ and $h$ do not change, the
accretion rate
\be
 \dot{M} = 2\pi \alpha h^2 r^2 \Omega_\m{K} \Sigma
 \label{eq:_accrrate}
\ee
 scales proportionally to $\Sigma$. This means that by varying
$\dot{M}$ at a constant $r_\m{m}$ we do not change the timescale
ratio $\tau_\m{g}/\tau_\m{m}$, i.e. the conclusions of Sect.
\ref{sect:_simulations} remain valid when the surface density is
decreased or increased at the inner edge of the disc.

As the planet recedes from the inner edge of the disc, its tidal
effect on the disc approaches the linear regime in which the
density perturbation $\Delta\Sigma$ is small compared to the
unperturbed density $\Sigma_0$. Eventually, $T$ should converge to
the linear torque which scales proportionally to $aM_\m{p}^2$
(Ward 1997), and Fig. \ref{torq_d1} shows that it is indeed the
case. Based on a particle approximation, the same scaling was
found by Lin and Papaloizou (1978) for the case of a small-mass
component of a binary with an external (i.e. circumbinary) disc.
(Strictly speaking, the formula given by Ward concerns the {\it
net torque} on the planet, but the torque components originating
exterior and interior to the orbit scale in the same way). After
some algebra we see that timescale ratios of planets with masses
$M_\m{p1}$ and $M_\m{p}$ which begin to spiral into cavities whose
edges are located, correspondingly, at $r_\m{m1}$ and $r_\m{m}$,
are related by the formula
\be
 \frac{\tau_\m{g1}}{\tau_\m{m1}}=
 \left(\frac{M_\m{p}}{M_\m{p1}}\right)^2\sqrt{\frac{r_\m{m}}{r_\m{m1}}}
 \,\frac{\tau_\m{g}}{\tau_\m{m}}.
 \label{eq:_ogle56}
\ee

\begin{figure}[!t]
 \includegraphics{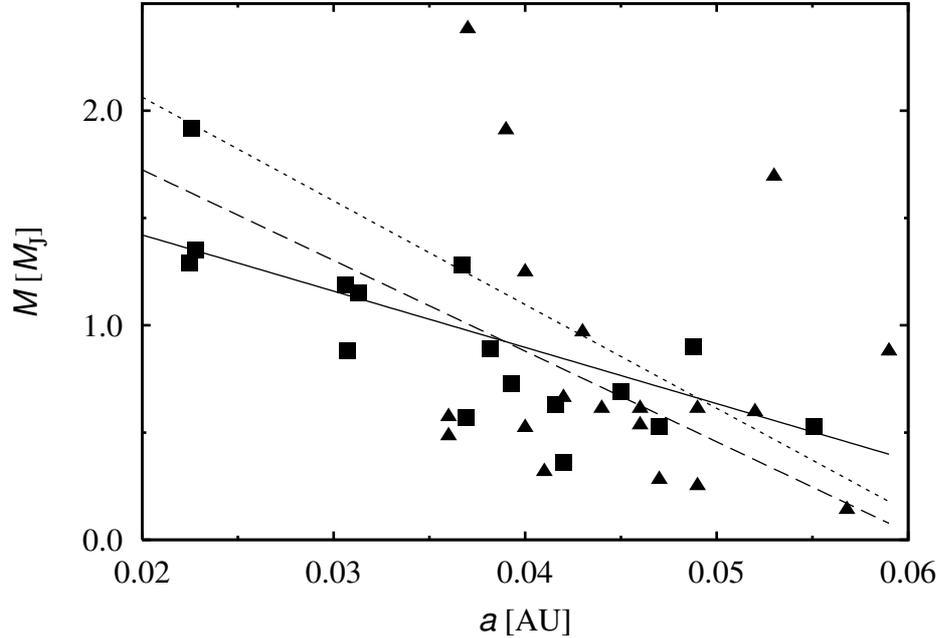}
 \caption {The mass - semimajor axis relation for the same planets
 as in Fig. \ref{fig: m-P_diagram}. Solid, dotted, and dashed
 line: fits to, correspondingly, all SPP known at the end of May 2007,
 final $(a,M_\m{p})$ distribution for $dN(M_\m{p})\sim
 M_\m{p}^{-1}dM_\m{p}$, final $(a,M_\m{p})$ distribution for
 $dN(M_\m{p})\sim M_\m{p}^{-2}dM_\m{p}$.}
 \label{fig: m-a_diagram}
\end{figure}

 Since we have found that $\sim 0.5 M_\m{J}$-planets are hardly
able to detach from the edge of the cavity, we may assume based on
Fig. \ref{fig: m-a_diagram} that the radii of the smallest
cavities were not much different from $\sim0.036$ AU. Let us
consider two identical planets, the first one spiralling into our
"standard" magnetosphere with $r_\m{m}=0.056$ AU, and the second
one into the "compact" magnetosphere with $r_\m{m1}=0.036$ AU.
According to equation (\ref{eq:_ogle56}), for the second planet
the ratio $\tau_\m{g}/\tau_\m{m}$ is just by $\sim25$\% larger
than for the first one. As a result, both planets reach a similar
stopping distance $d_\m{s}$ from the edge of the disc, and to a
first approximation we may assume that the function
$d_\m{s}(M_\m{p},r_\m{m})$ does not depend on $r_\m{m}$.

According to Marcy et al. (2005), the distribution of planetary
masses is affected very little by the unknown inclination of
orbits, and based on available data, one may adopt
\be
 dN(M_\m{p})\sim M_\m{p}^{-1}dM_\m{p}\,.
 \label{eq:_mhist}
\ee
 As we have mentioned before, the distribution
of magnetospheric radii is unknown, but since there is no obvious
reason to favour any value of $r_\m{m}$ we may assume that all
values between $r_\m{m1}$ and $r_\m{m}$ are equally probable. With
these assumptions, the joint distribution of $M_\m{p}$ and
$r_\m{m}$ is given by
\be
 F(M_\m{p},r_\m{m})={\cal N}\,M_\m{p}^{-1}\,\chi_1(M_\m{p})\chi_2(r_\m{m}),
\ee
 where $\chi_1(x)=1$ for $0.5\le x \le 2.5$ and 0 otherwise (2.5 $M_\m{J}$
is the estimated mass of the most massive SPP in Fig. \ref{fig:
m-a_diagram}), $\chi_2(x)=1$ for $0.036\le x \le 0.056$ and 0
otherwise, whereas
\be
 {\cal N} = \left(\int_{0}^{\infty}\int_{0}^{\infty}
 m^{-1}\,\chi_1(m)\,\chi_2(x)\,\,dx\,\,dm
 \right)^{-1}.
\ee
 The ratio $\tau_\m{g}/\tau_\m{m}$ decreases rather sharply
with the increasing mass of the planet, but in order to remain on
the conservative side we may adopt that the stopping distance
scales linearly with $M_\m{p}$, according to the formula
\be
 \frac{d_\m{s}}{\m{AU}}=0.012 \frac{M_\m{p}}{M_\m{J}}-0.006\,,
 \label{eq:_stopd}
\ee
 which in agreement with our results yields $d_\m{s}=0$ AU for
$M_\m{p}=0.5 M_\m{J}$ and $d_\m{s}=0.006$ AU for $M_\m{p}=1
M_\m{J}$.

Let us again imagine an evolving population of planets represented
by points scattered on the $(a,M_\m{p})$ plane. Every planet
begins to spiral down from its $r_\m{m}$, and stops at
$a_\m{s}=r_\m{m}-d_\m{s}$, with $d_\m{s}$ depending primarily on
the planet's mass. Thus, the final distribution of points is
skewed toward smaller $a$, i.e. a correlation betwen $M_\m{p}$ and
$a$ is generated. The least-square linear fit to the final
distribution
\be
 M_\m{p}=\alpha*a +\beta
 \label{eq:_lfit}
\ee
 is given by the standard formulae
\begin{eqnarray}
 \alpha&=&\frac{
 \int_{0}^{\infty}\int_{0}^{\infty}\left(m-\bar{M}_\m{p}\right)
 *\left[\left(x-d_\m{s}(m)\right)-\left(\bar{r}_\m{m}-\bar{d}_\m{s}\right)
 \right] F(m,x)\,\,dx\,\,dm}
 {\int_{0}^{\infty}\int_{0}^{\infty}
 \left[\left(x-d_\m{s}(m)\right)-\left(\bar{r}_\m{m}-\bar{d}_\m{s}\right)
 \right]^2 F(m,x)\,\,dx\,\,dm}\nonumber \\
 \beta&=&\bar{M}_\m{p}-\alpha*\left(\bar{r}_\m{m}-\bar{d}_\m{s}\right),
 \label{eq:_beta}
\end{eqnarray}
 and is shown in Fig. \ref{fig: m-a_diagram}. For comparison,
analogous fit obtained for a steeper mass distribution
$dN(M_\m{p})\sim M_\m{p}^{-2}dM_\m{p}$ is also plotted. Given the
simplicity of our approach, the qualitative agreement of
theoretical fits with the observational one looks quite
encouraging.

So far we have neglected tidal interactions between the planet and
the star. Following LBR and recent theoretical as well as
observational evidence (Bouvier et al. 2006; Herbst \& Mundt 2005,
and references therein) we assume that T Tauri stars are
rotationally locked to inner edges of their accretion discs. Since
a locked star rotates at
$\Omega_\star\approx\Omega(r_\m{m})<\Omega_\m{p}$, the resulting
stellar torque onto the planet is negative, causing orbital radius
of the planet to evolve according to the formula
\be
 a(t)^{6.5}=a(0)^{6.5}-\frac{13}{2}
  \frac{3k_2\star}{Q_\star}\frac{M_\m{p}}{M_\star}R_\star^5\sqrt{\m{G}M_\star}\,t\,;
  \label{eq:star_tide1}
\ee
  see e.g. P\"atzold et al. (2004). For $M_\star = M_\odot$ we have
\be
 x(t)^{6.5}= x_0^{6.5}-3.2\times10^{-6}\frac{M_\m{p}}{M_\m{J}}
 \left(\frac{R_\star}{R_\odot}\right)^{5}t\,,
 \label{eq:star_tide2}
\ee
 where $x\equiv a/R_\odot$, $t$ is expressed in years, and a value
of $1.2\times10^8$ was adopted for $Q_\star/k_2$ (P\"atzold et al.
2004). Under the influence of stellar tides alone a 1 $M_\m{J}$
planet departing from the edge of the magnetosphere at
$r_\m{m}=0.056$ AU would spiral down to the surface of a
$3R_\odot$ star within $\sim1.3\times10^{10}$ yr, whereas for the
same planet departing from $r_\m{m1}=0.036$ AU the spiralling
would take $\sim7.5\times10^{8}$ yr. In fact, the spiralling time
would be even longer because the star contracts, causing the
stellar torque on the planet to decrease. According to Allain
(1998), within $\sim10^7$ yr $R_\star$ can easily shrink by a
factor of 5 if the star does not accrete, and when accretion is
allowed for the contraction proceeds even more quickly (Siess et
al. 1999). As a result, it is highly unlikely for stellar tides to
significantly influence the orbital evolution of the planet within
the disc lifetime.

On the other hand, equation (\ref{eq:star_tide1}) predicts a
faster evolution for more massive planets, and assuming that
$\Omega_\star$ remains smaller than $\Omega_\m{p}$ one might
expect that Fig. \ref{fig: m-P_diagram} is just a snapshot of a
distribution undergoing slow secular changes due to stellar tides.
However, at the end of the pre--Main Sequence phase the radius of
the star is equal to $\sim R_\odot$, and the stellar torque on the
planet strongly decreases. As a result, the spiraling timescale of
a 1 $M_\m{J}$ planet from $r_\m{m1}=0.036$ AU lengthens to well
above $10^{10}$ yr.

In the final note we would like to stress that we do not claim
that the problem of correlation between SPP masses and periods has
been solved. We merely report a mechanism which may qualitatively
account for the observed trend. An alternative mechanism was
disussed by Faber, Rasio \& Willems (2005) and Rasio \& Ford
(2006), who argue that SPP are likely to have originated from the
tidal capture of planets on originally highly eccentric orbits. It
is also possible that both mechanisms have been at work; provided,
of course, that the correlation is real and will be confirmed by
future observations.
\section*{Acknowledgments}
This work was supported through grant 1P03D 02626 from the Polish
Ministry of Science and by the European Research Training Network
"The Origin of Planetary Systems" (contract number
HPRN-CT-2002-00308). The software used in this work was in part
developed by the DOE-supported ASC / Alliance Center for
Astrophysical Thermonuclear Flashes at the University of Chicago.
The simulations were partly performed at the Interdisciplinary
Centre for Mathematical and Computational Modeling in Warsaw.

\section*{Appendix: Code testing}
Before the version of FLASH adapted to the problem of disc --
planet interaction was used for the proper simulations, we had
tested it for compatibility with analytical expressions for the
disc torque on the planet obtained by Ward~(1997).

Consider a thin disc with a surface density $\sigma(r,\phi)$,
orbiting a central star of mass $M_\star$. The mass of the disc is
minute compared to $M_\star$, and its self-gravity is negligible,
causing orbital velocity and epicyclic frequency of the disc
matter, $\Omega$ and $\kappa$, to be nearly Keplerian. Assume that
the temperature distribution in the disc does not depend on time,
with the local sound speed $c_\m{s}$ fixed according to
(\ref{eq:sounds}). The disc is perturbed by a planet of mass
$M_\m{p} = \mu M_\star$, which moves around the star on a circular
orbit of radius $a$. As the perturbations are nonaxisymmetric, the
planet is subject to a gravitational torque $T$ from the disc.
According to Ward~(1997), the torque density ${\rm d}T/{\rm d}r$
is given by
\begin{equation}
   \frac{{\rm d}T}{{\rm d}r} = \epsilon \frac{2\mu^2(\sigma a^2)(a\Omega_\m{p})^2
   m_\m{r}^4 \psi^2}{r(1+4\xi^2)} \left(\frac{\Omega_\m{p}}{\kappa}\right)^2
   \ ,
\end{equation}
where $\Omega_\m{p}$ is the angular orbital velocity of the
planet, $\xi = m_\m{r}c_s/r\kappa$, and $\epsilon$ is equal to
$+1$ ($-1$) for torques originating in the disc interior
(exterior) to the planet's orbit. The function $m_\m{r}$ is given
by the formula
\begin{equation}
   m_\m{r} = \sqrt{\frac{\kappa^2}{(\Omega-\Omega_\m{p})^2 - c_\m{s}^2/r^2}} =
   \left[\left(1-\sqrt{\frac{r^3}{a^3}(1+h^2)}\right)^{2}-h^2\right]^{-1/2}\ ,
\end{equation}
where $\Omega^2 = \Omega_\mathrm{K}^2 - c_\m{s}^2/r^2$ and
(\ref{eq:sounds}) were used; whereas the function $\psi$ -- by the
formula
\begin{equation}
   \psi = \frac{\pi}{2}\left[\frac{1}{m_\m{r}}\left|\frac{{\rm d}b_{1/2}^m(x)}
   {{\rm d} x} \right| + 2\sqrt{1+\xi^2}\:b_{1/2}^m(x) \right] \ ,
\end{equation}
where $b_{1/2}^m(x)$ is the Laplace coefficient with argument
$x=r/a$.

\begin{figure}[!h]
\includegraphics{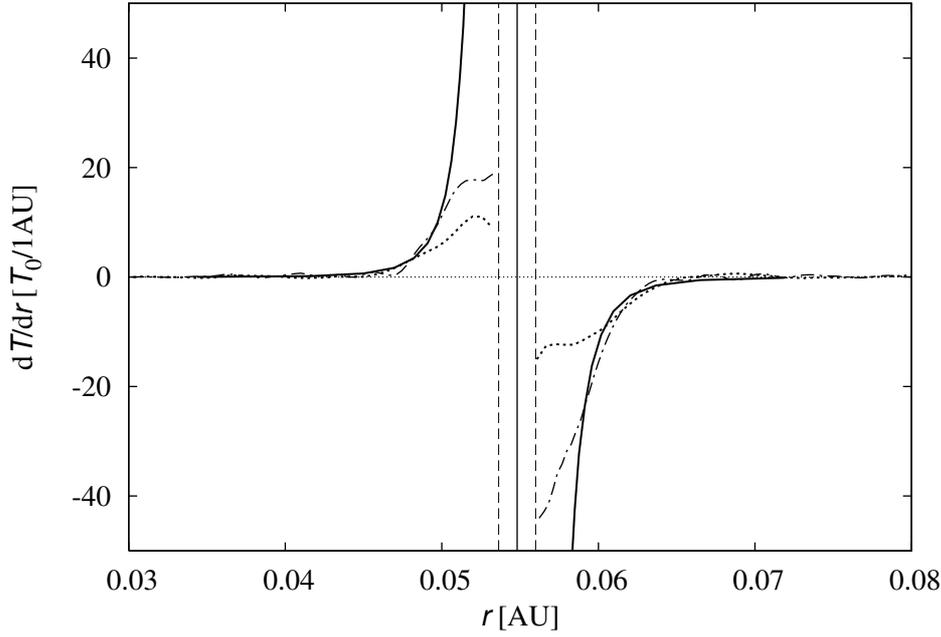}
  \caption
  {Torque densities obtained numerically for $n_r\times n_\phi$ =
  256$\times$128 points (dotted) and
  512$\times$256 points (dash-dotted),
  compared with the analytical solution (solid). Vertical lines:
  location of the planet (solid) and the torque cuttoff limits at the Roche
  radius (dashed). $T_0$ is the normalized torque defined as $T_0 =
  \uppi\mu^2\sigma a^2(a\Omega_\m{p})^2(a/h)^3$. The plotted
  values are averages from 100 time steps (roughly one radian of
  the orbit).
  }
  \label{dTdr}
\end{figure}
To calculate the torque density from the numerical models, we
integrate contributions from all grid cells forming a thin annulus
at a distance $r$ from the star\,:
\begin{equation}
   \frac{{\rm d}T}{{\rm d}r} = \m{G}\int_0^{2\pi} \frac{M_\m{p}\:
    r\sin(\phi-\phi_\m{p})}
    {d^2+\varepsilon^2r_\m{R}^2}\:
    \sigma(r,\phi) r {\rm d}\phi ,
\end{equation}
where
\begin{equation}
    d = \sqrt{{a}^2 + r^2 -
    r\cos(\phi -\phi_\mathrm{p})}
\end{equation}
is the distance between the planet and the point $(r,\phi)$ on the
annulus; $\phi_\mathrm{p}$ stands for the positional angle of the
planet, $r_\mathrm{R}$ is the Roche radius, and $\varepsilon$ is a
dimensionless softening parameter.

 For the test runs we adopted $M_\m{p} = 10\,M_\m{E}$ (Earth
mass) and $a = r_\m{m}$. The calculations were performed on a
polar grid $(r,\phi)$ extending radially from 0.03 to 0.18~AU. The
disc had a constant surface density $\sigma = 2.28\times
10^5$~g\,cm$^{-2}$, so that the corresponding mass of the disc
matter contained within the grid was equal to $2.5\,M_\m{J}$
(Jupiter mass). The softening parameter $\varepsilon$ was set to
0.6. No explicit viscosity was used, and no special treatment was
applied to the matter within the Roche lobe of the planet. The
disc gas could flow freely through the radial boundaries of the
grid.

We ran two test simulations with different resolutions. The
results are shown in Fig. \ref{dTdr}, from which it is evident
that on the finer grid ($512\times256$ points) the code is able to
calculate reasonably accurate torques from the parts of the disc
that are more distant than $\sim0.005$~AU or $\sim10$ grid cells
from the planet. Since that area contributes less than 20\% of the
total torque from the disc, it may seem rather discouraging. The
obvious conclusion is that if the planet is not able to open a
gap, a really very high resolution is necessary to follow its
migration through the disc. Obviously, for gap-opening planets the
resolution requirements are much less stringent.

\end{document}